\documentclass[12pt,epsfig]{iopart}
\usepackage{epsfig}
\begin{document}

\title[Wroblewski Parameter \& Lattice QCD]{The Wroblewski Parameter from Lattice QCD
\vskip-1.4cm\hfill\small hep-ph/0403172, TIFR/TH/04-06\vskip1.1cm
}

\author{Rajiv V. Gavai\footnote[3]{Speaker at the conference}
and Sourendu Gupta}

\address{Department of Theoretical Physics, 
Tata Institute of Fundamental Research, \\
Homi Bhabha Road, Mumbai 400 005, India}

\begin{abstract}
Enhancement of strangeness production has since long been proposed as a
promising signal of quark-gluon plasma production.  A convenient indicator for
it is the Wroblewski parameter which has been shown to be about a factor two
higher in heavy ion collisions.   Using a method proposed by us earlier, we
obtained lattice QCD results for the Wroblewski parameter from our simulations 
of QCD with two light quarks both below and above the chiral transition.  Our
first principles based and parameter free result compare well with the A-A data. 

\end{abstract}



\maketitle

\section{Introduction}
Amongst the many proposed signatures of quark-gluon plasma in relativistic
heavy ion collisions, enhancement of strangeness production  \cite{RM} is a
promising signal.    Proposed in the very early stages of this field, it was
based on simple model considerations like many other signals.  Exploiting the
fact, that the strange quark mass  is smaller than the expected transition
temperature whereas the mass of the lowest strange hadron is significantly
larger, it was argued that the production rate for strangeness in the QGP
phase, $\sigma_{QGP} (s \bar s)$ is greater than that in the hadron gas phase,
$ \sigma_{HG}(s \bar s)$.  While this energy threshold argument for strangeness
production in the two phases is qualitatively appealing, one has to face
quantitative questions of details for any meaningful comparison with the data.
Since the temperature of the plasma produced in RHIC, or even LHC, may not be
sufficiently high for perturbative QCD to be applicable, estimation of
strangeness production by lowest order processes like $gg \to s \bar s$ could
be misleading.  Indeed, it is now well-known that even for charm production,
the next order correction to $gg \to c \bar c$ is equally large.

A variety of aspects of the strangeness enhancement have been studied and 
many different variations have been proposed.   One very useful way of
looking for strangeness enhancement is the Wroblewski parameter \cite{Wr}.
Defined as the ratio of newly created strange quarks to light quarks, 
\begin{equation}
\lambda_s = \frac{2\langle s\bar s\rangle}{\langle u\bar u+d\bar d\rangle}
\end{equation}
the Wroblewski parameter has been estimated for many processes using a hadron
gas fireball model \cite{BH}.  An interesting finding from these analyses is
that $\lambda_s$ is around 0.2 in most processes, including proton-proton
scattering, but is about a factor of two higher in heavy ion collisions. An
obvious question one can ask is whether this rise by a factor of two can be
attributed to the strangeness enhancement due to quark gluon plasma and if yes,
whether this can be quantitatively demonstrated.  We show below how quark
number susceptibilities, obtained from simulations of lattice QCD, may be
useful in answering these questions.


\section{$\lambda_s$ from Quark Number Susceptibilities}


Quark number susceptibilities (QNS) can be calculated from first principles
using the lattice formulation. Assuming three flavours, $u$, $d$, and $s$
quarks, and denoting by $\mu_f$ the corresponding chemical potentials, the QCD
partition function is
\begin{equation}
{\cal Z} =  \int D U~~ \exp(-S_G)  \prod_{f=u,d,s} 
{\rm Det}~M(m_f, \mu_f)~~.
\label{zqcd}
\end{equation}
Note that the quark mass and the corresponding chemical potential enter only
through the Dirac matrix $M$ for each flavour.  Defining $\mu_0 = \mu_u + \mu_d
+ \mu_s$ and $\mu_3 = \mu_u - \mu_d$, the baryon and isospin densities and the
corresponding susceptibilities can be obtained as:  

\begin{equation}
\qquad \qquad n_i =  \frac{T}{V} {{\partial \ln {\cal Z}}\over{\partial \mu_i}}, \qquad
\chi_{ij} =  \frac{T}{V} {{\partial^2 \ln {\cal Z}}\over{\partial \mu_i 
\partial \mu_j} }  ~~.
\label{nchi}
\end{equation}

QNS in \eref{nchi} are crucial for many quark-gluon plasma signatures which are
based on fluctuations in globally conserved quantities such as baryon number or
electric charge.  Theoretically, they serve as an important independent check
on the methods and/or models which aim to explain the large deviations of the
lattice results for pressure $P$($\mu$=0) from the corresponding perturbative
expansion.  Here we will be concerned with the Wroblewski parameter 
which we \cite{us} have argued can be estimated from the quark number 
susceptibilities:
\begin{equation}
\lambda_s = {2 \chi_s \over { \chi_u +\chi_d}}~. 
\label{wrob}
\end{equation}

In order to use \eref{wrob} to obtain an estimate for comparison with
experiments, one needs to compute the corresponding quark number
susceptibilities on the lattice first and then take the continuum limit.  All
susceptibilities can be written as traces of products of $M^{-1}$ and various
derivatives of $M$ with respect to $\mu$.  With $m_u = m_d$, diagonal 
$\chi_{ii}$'s can be written as

\begin{eqnarray}
\label{chiexp1}
\chi_0 &=& \frac{T}{2V} [ \langle {\cal O}_2(m_u) + \frac{1}{2} {\cal O}_{11}(m_u) \rangle ] \\ 
\label{chiexp2}
\chi_3 &=& \frac{T}{2V} ~~ \langle {\cal O}_2(m_u) \rangle  \\ 
\label{chiexp3}
\chi_s &=& \frac{T}{4V} [ \langle {\cal O}_2(m_s) + \frac{1}{4} {\cal O}_{11}(m_s) \rangle ] ~~.
\end{eqnarray} 

\noindent
Here ${\cal O}_2 = {\rm Tr}~M^{-1}_u M_u'' - {\rm Tr} ~M^{-1}_u M_u'M^{-1}_u
M_u'$, and $ {\cal O}_{11}(m_u) = ({\rm Tr}~M^{-1}_u M_u' )^2$.  The traces are
estimated by a stochastic method: $ {\rm Tr}~A = \sum^{N_v}_{i=1} R_i^\dag A
R_i / 2N_v$, and $ ({\rm Tr}~A)^2 = 2 \sum^{L}_{i>j=1} ({\rm Tr}~A)_i ({\rm
Tr}~A)_j/ L(L-1)$, where $R_i$ is a complex vector from a set of $N_v$,
subdivided further in L independent sets.

\begin{figure}[htb]
\vspace{-0.2cm}
\begin{minipage}{0.49\textwidth}
\epsfxsize=6.4cm
\epsfbox{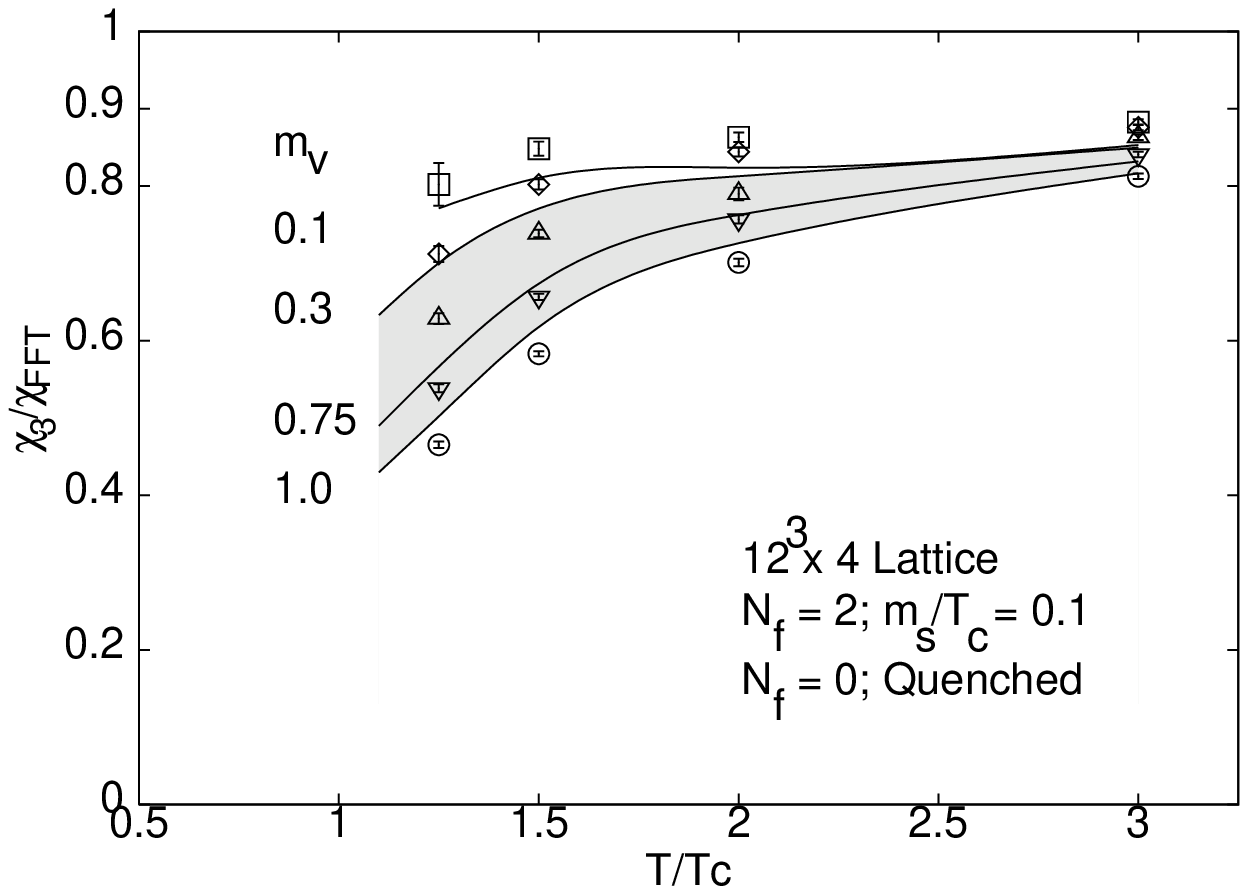}
\end{minipage}
\begin{minipage}{0.49\textwidth}\raggedright
\epsfxsize=6.4cm
\epsfbox{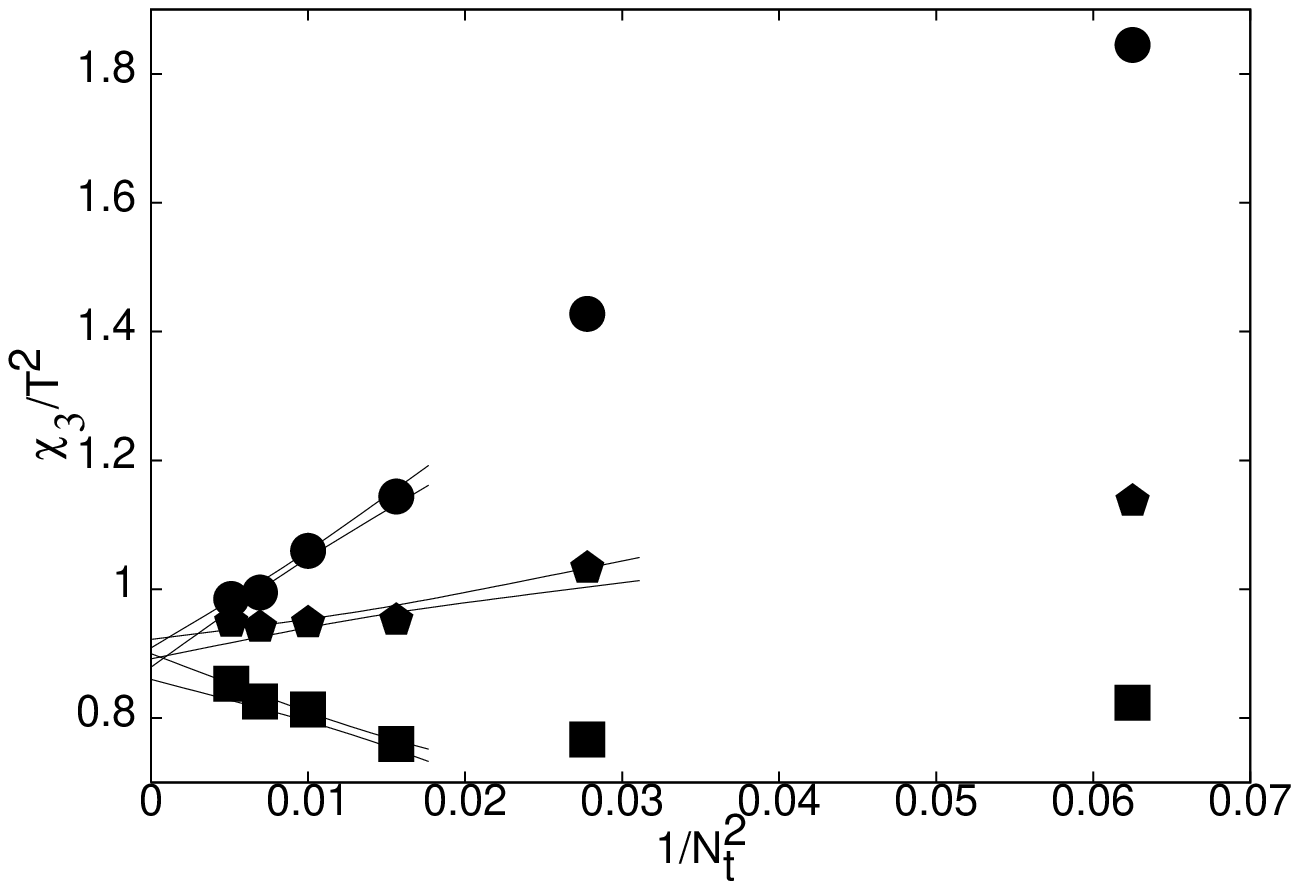}
\end{minipage}
\vspace{-0.3cm}
\label{fig1}
\caption{Comparison of quenched and full QCD (left) results and typical 
continuum extrapolation results (right).}
\vspace{-0.3cm}
\end{figure}


Left panel of Figure 1 displays results \cite{us1} for the susceptibilities as
a function of temperature in units of $T_c$, where $T_c$ is the transition
temperature.  Normalized to the corresponding ideal gas results on the same
lattice, i.e, the infinite temperature limit, results for QCD with two light
dynamical quarks of mass $0.1~T_c$ are shown as points whereas the continuous
curves correspond to the results in the quenched approximation.  The valence
quark mass $m_{\rm v}$, appearing in \eref{chiexp1}-\eref{chiexp3}, is shown in
the figure in units of $T_c$.  Note that $T_c$ in these two cases differ by a
factor of 1.6 but the results for the corresponding susceptibilities as a
function of $T/T_c$ differ by a few per cent only.  Encouraged by this, we
investigated the continuum limit for the quenched case by increasing the
temporal lattice size from 4 to 14, as shown in the right panel of Figure 1 for
$T = 2~T_c$.  The continuum results for QNS thus obtained in the quenched
approximation are exhibited in Figure 2 for small $m_{\rm v}$.

\begin{figure}[htb]
\vspace{-0.2cm}
\begin{minipage}{0.49\textwidth}
\epsfxsize=6.4cm
\epsfbox{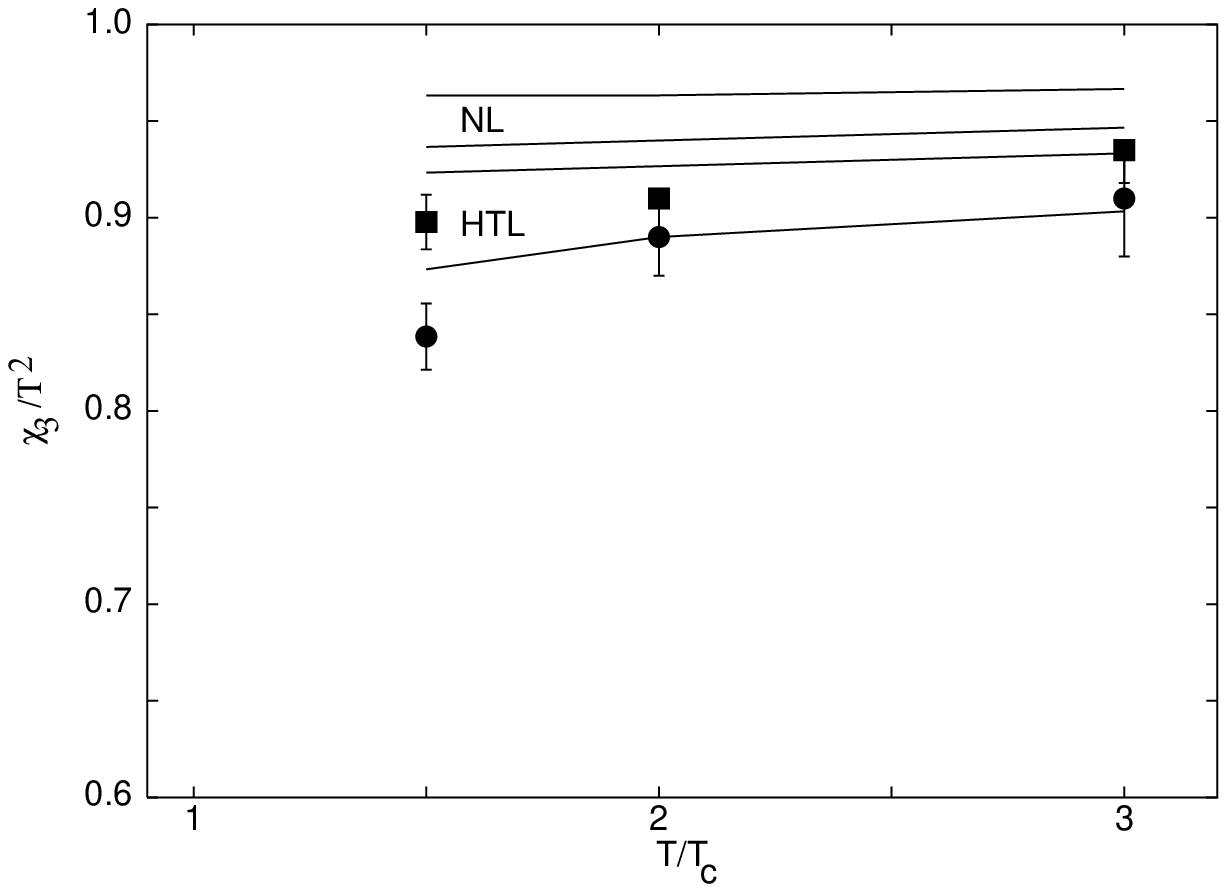}
\end{minipage}
\begin{minipage}{0.49\textwidth}\raggedright
\epsfxsize=6.4cm
\epsfbox{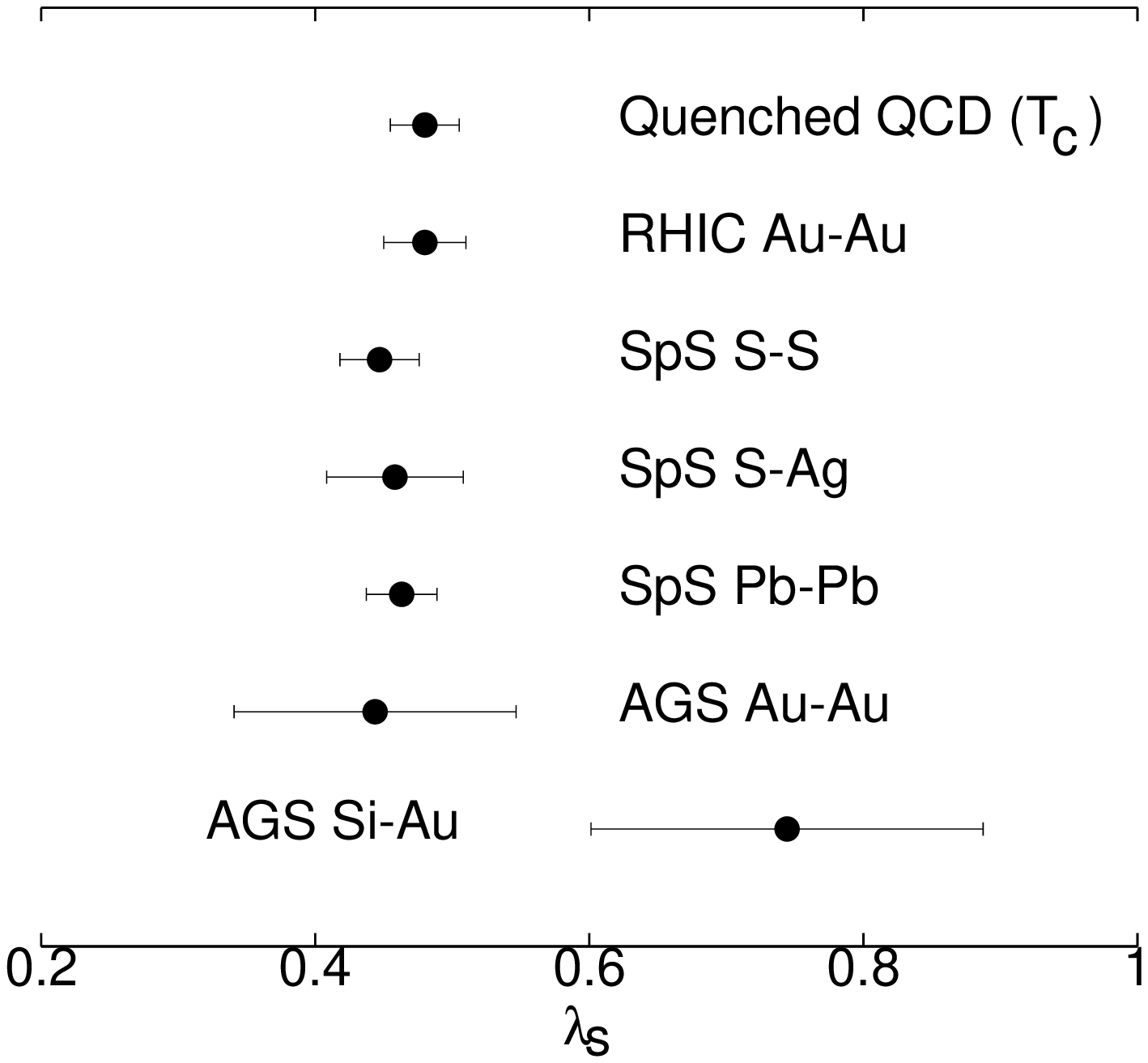}
\end{minipage}
\vspace{-0.3cm}
\label{fig2}
\caption{Quenched QCD results in continuum limit (left) and comparison of
the corresponding $\lambda_s$ with RHIC and SPS experiments (right).}
\vspace{-0.3cm}
\end{figure}


The strange quark susceptibility was obtained from the same simulations by
simply choosing $m_{\rm v}/T_c = 1$ (in both full and quenched QCD in view of
Figure 1). Using \eref{wrob}, $\lambda_s(T)$ can then be easily obtained, and
extrapolated to $T_c$ by employing simple ans\"atze.  The resultant
$\lambda_s(T_c)$ in quenched QCD is shown in the right panel of Figure 2 along
with the results obtained from the analysis of the RHIC and SPS data in the
fireball model\cite{BH}.  The systematic error coming from extrapolation is not
shown but is of approximately the same size as the shown statistical error.
The agreement of the lattice results with those from RHIC and SPS is indeed
very impressive.

\begin{figure}[htbp]\begin{center}
 
\vspace{-0.2cm}
\epsfig{height=7cm,file=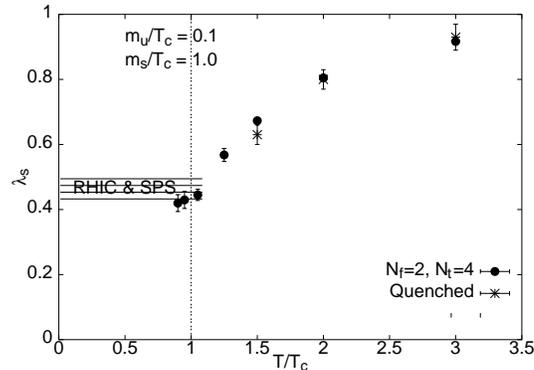, angle=270}
 
\vspace{-0.3cm}
\caption{$\lambda_s$ as a function of temperature for full and quenched QCD.}
 
\vspace{-0.3cm}
\label{fig3}\end{center}
\end{figure}

The nice agreement needs to be treated cautiously, however, in view of the
various approximations made. Let me list them in order of severity.

\begin{itemize}

\item The result is based on quenched QCD simulations and extrapolation to
$T_c$.  As seen from Figure 1, the quark number susceptibilities, and hence
$\lambda_s(T)$, are expected to change by only a few per cent.  Since the
nature of the phase transition does depend strongly on the number of dynamical
quarks, a direct computation near $T_c$ for full QCD is desirable.  We  are
currently making such a computation and have some preliminary results for full
QCD with two light dynamical quarks for lattices with four sites in temporal
direction.  These are shown in Figure 3 along with the continuum quenched
results for $\lambda_s(T)$ and the band for experimental results.  While the
emerging trend is encouraging, further exploration with varying strange quark
mass, temporal lattice size (to obtain continuum results) and spatial volume is
still necessary. 

\item The experiments at RHIC and SPS have nonzero albeit small $\mu$ whereas
the above result used $\mu=0$. Based on both lattice QCD and fireball model
considerations, $\lambda_s$ is expected to change very slowly for 
small $\mu$.  This can, and should, be checked by direct simulations.

\item Lattice simulations yield real quark number susceptibility whereas for
particle production its imaginary counterpart is needed.  Assuming that the
characteristic time scale of plasma are far from the energy scales of strange
or light quark production, one can relate \cite{book} the two to justify
the use of lattice results in obtaining $\lambda_s$.   Observation of spikes 
in photon production may falsify this assumption. 

\end{itemize}
\section{Summary}

Quark number susceptibilities which can be obtained from first principles
using lattice QCD contribute substantially to the physics of RHIC signals.
In particular, the continuum limit of $\chi_u$ and $\chi_s$, obtained
in quenched QCD, leads to $\lambda_s(T)$. Its extrapolation to $T_c$ appears
to be in good agreement with results from SPS and RHIC. First full QCD
results near $T_c$ confirm this as well, although many technical issues,
e.g, finite lattice cut-off or strange quark mass, need to be sorted out still.

\vspace{0.5cm}


\end{document}